\documentclass[
aps,
nofootinbib,
prd,
superscriptaddress,
tightenlines,
notitlepage,
twocolumn,
showpacs,
floatfix
]{revtex4-1}

\usepackage[T1]{fontenc}
\usepackage{amssymb}
\usepackage{graphicx}
\usepackage{amsmath}
\usepackage{xcolor}
\usepackage{tensor}
\usepackage{epstopdf}
\usepackage{multirow}
\usepackage{extarrows}
\usepackage{float}
\usepackage{graphicx,subfigure}
\usepackage{braket}
 \usepackage[normalem]{ulem}
\usepackage[colorlinks=true,
citecolor=blue,
linkcolor=red,
anchorcolor=black,
urlcolor=purple]{hyperref}

\def\al{\alpha}
\def\be{\beta}

\def\de{\delta}

\def\ka{\kappa}
\def\la{\lambda}

\def\mn{{\mu\nu}}

\def\lsim{\mathrel{\rlap{\lower4pt\hbox{\hskip1pt$\sim$}}
    \raise1pt\hbox{$<$}}}
\def\gsim{\mathrel{\rlap{\lower4pt\hbox{\hskip1pt$\sim$}}
    \raise1pt\hbox{$>$}}}
\def\sqr#1#2{{\vcenter{\vbox{\hrule height.#2pt
         \hbox{\vrule width.#2pt height#1pt \kern#1pt
         \vrule width.#2pt}
         \hrule height.#2pt}}}}

\def\prt{\partial}
\def\lrpartial{\raise 1pt\hbox{$\stackrel\leftrightarrow\partial$}}

\def\part2{\partial_\alpha \partial^\alpha}

\def\pt#1{\phantom{#1}}

\def\xx'{|\vec x -\vec x'|}

\def\b2{b^\al b_\al}

\newcommand{\beq}{\begin{equation}}
\newcommand{\eeq}{\end{equation}}
\newcommand{\bea}{\begin{eqnarray}}
\newcommand{\eea}{\end{eqnarray}}
\newcommand{\bit}{\begin{itemize}}
\newcommand{\eit}{\end{itemize}}
\newcommand{\rf}[1]{(\ref{#1})}
\newcommand\bw{\begin{widetext}}
\newcommand\ew{\end{widetext}}
\newcommand{\td}{\mathrm{d}}

\newcommand{\nn}{\nonumber}

\def\0{{\sst{(0)}}}
\def\1{{\sst{(1)}}}
\def\2{{\sst{(2)}}}
\def\3{{\sst{(3)}}}
\def\4{{\sst{(4)}}}
\def\5{{\sst{(5)}}}
\def\6{{\sst{(6)}}}
\def\7{{\sst{(7)}}}
\def\8{{\sst{(8)}}}
\def\sst#1{{\scriptscriptstyle #1}}

\begin{document}

\title{\boldmath Extended thermodynamics of the bumblebee black holes}

\author{Zhan-Feng Mai} \email{zhanfeng.mai@gmail.com}
\affiliation{Kavli Institute for Astronomy and Astrophysics, Peking University,
Beijing 100871, China}

\author{Rui Xu} \email{xuru@pku.edu.cn}
\affiliation{Kavli Institute for Astronomy and Astrophysics, Peking University,
Beijing 100871, China}
\affiliation{Department of Astronomy, Tsinghua University, Beijing, 100084, China}

\author{Dicong Liang}

\affiliation{Kavli Institute for Astronomy and Astrophysics, Peking University,
Beijing 100871, China}

\author{Lijing Shao}

\affiliation{Kavli Institute for Astronomy and Astrophysics, Peking University,
Beijing 100871, China}
\affiliation{National Astronomical Observatories, Chinese Academy of Sciences,
Beijing 100012, China}

\begin{abstract}
As a vector-tensor theory including nonminimal coupling between the Ricci tensor
and a vector field, the bumblebee gravity is a potential theory to test Lorentz
symmetry violation. Recently, a new class of numerical spherical black holes in
the bumblebee theory was constructed. In this paper, we investigate the
associated local thermodynamic properties. By introducing a pair of conjugated
thermodynamic quantities $X$ and $Y$, which can be interpreted as an extension
of electric potential and charge of the Reissner-Nordstr\"om black holes, we
numerically construct a new first law of thermodynamics for bumblebee black
holes. We then study the constant-$Y$ processes in the entropy-charge parameter
space. For the constant-$Y$ processes, we also calculate the heat capacity to
study the local thermodynamic stability of the bumblebee black holes. For a
negative nonminimal coupling coefficient $\xi$, we find both divergent and
smooth phase transitions.  For a positive but small $\xi$, only a divergent
phase transition is found. It turns out that there is a critical value
$0.4\kappa <\xi_c < 0.5\kappa$ such that when $\xi_c < \xi<2\kappa$, even
the divergent phase transition disappears and the bumblebee black holes thus
become locally thermodynamically unstable regardless of the bumblebee charge. As
for $\xi>2\kappa$, the smooth phase transition arises again but there no longer
exists any discontinuous phase transition for the bumblebee black holes. 
\end{abstract}

\maketitle
\flushbottom

\noindent

\section{Introduction}

The bumblebee theory is a vector-tensor theory as well as an extension of the
Einstein-Maxwell theory. In the bumblebee theory, the vector field $B_\mu$,
which is also called the bumblebee field, is nonminimally coupled with the Ricci
tensor quadratically and has a nonzero background value leading to spontaneous
Lorentz symmetry violation by minimizing its potential term $V$. The action
associated with the bumblebee theory is \cite{Kostelecky:2003fs}
\begin{eqnarray}\label{actionbum}
 I&=&\int \td^4 x \sqrt{-g} \Big( \frac{1}{2\kappa} R+ \frac{\xi}{2\kappa}B^\mu B^\nu R_{\mu\nu} -\frac{1}{4}B^{\mu\nu}B_{\mu\nu}  \, \cr
~\cr
&& -V(B^\mu B_\mu \pm b^2) \Big)+ S_m\, ,
\end{eqnarray}
where $\kappa \equiv 8\pi$, $B_\mu$ denotes the bumblebee field and its field
strength is defined as $B_{\mu\nu} \equiv \partial_\mu B_\nu - \partial_\nu
B_\mu$. The coupling constant $\xi$ indicates the strength of the quadratic
interaction between the bumblebee field and the Ricci tensor $R_{\mu\nu}$. The
last term $S_m$ denotes the action given by normal matter. Furthermore, $V(B^\mu
B_\mu \pm b^2)$ refers to a cosmological potential. In general, for obtaining a
stable vacuum of spacetime, the potential $V$ will be minimized if $B^\mu B_\mu
= \mp b^2$. In other words, in the background of $B^\mu B_\mu = \mp b^2$, there
exists a preferred direction of the spacetime so that the Lorentz symmetry is
broken.

If considering $V$ with a nonzero minimal value, this potential provides a
cosmological constant. However, for an unknown expression of the potential $V$,
a practical way to probe the Lorentz breaking is considering a constant
bumblebee background configuration in an asymptotically flat spacetime in the
case of vanishing minimal potential value. In the weak field limit, the constant
bumblebee field background is related to the coefficients associated with
Lorentz violation in the Standard Model extension (SME) framework
\cite{Kostelecky:2009zp, Kostelecky:2011gq}. In this paper, we consider the
nonzero bumblebee field background configuration, $B^\mu = b^\mu$ satisfying
$b^\mu b_\mu = \pm b^2$, which minimizes the potential to zero, namely
\begin{align}
\left.V(B^\mu B_\mu \pm b^2)  \right|_{B^\mu = b^\mu} &= 0 \,, \\
\left.V' (B^\mu B_\mu \pm  b^2)  \right|_{B^\mu = b^\mu} &= 0 \, ,
\end{align}
where $V'(x) \equiv \td V(x)/\td x$.

In addition to effects in the weak-field region, effects in the strong field
region of Lorentz symmetry breaking is an intriguing topic in the bumblebee
theory as well. Black holes, as the most compact objects in our universe, have
many important and intriguing properties in both theoretical study and
astrophysical observation.  Recently, several analytical solutions, including
spherical black holes as well as those with slow rotation, and their associated
properties in the bumblebee gravity have been investigated
\cite{Bertolami:2005bh, Oliveira:2021abg, Gullu:2020qzu, Maluf:2020kgf,
Izmailov:2022jon, Liu:2022dcn}. \citet{Bertolami:2005bh} constructed a class of
Schwarzschild-like black hole solutions analytically where the bumblebee field
has only a nonzero radial component.  Furthermore, another class of spherical
bumblebee black holes, involving the bumblebee hair only with time component,
has also been numerically constructed by \citet{Xu:2022frb}, and its extreme
mass ratio inspirals (EMRIs) for probing the bumblebee hair \cite{Liang:2022gdk}
as well as the possibility for Event Horizon Telescope (EHT) observation has
been studied \cite{Xu:2023xqh}. However, the thermodynamics of this class of
numerical black holes, as a very important theoretical aspect of this topic,
still needs to be addressed. 

Black hole thermodynamics provides an enlightening way to study various models
of gravity \cite{PhysRevD.13.191, PhysRevD.7.2333, Hawking:1975vcx,
Bardeen:1973gs, Davies:1977bgr}. Interpreting black holes as thermodynamic
systems, one can investigate black holes by only a few thermodynamic quantities.
Additionally, the thermodynamic quantities of black holes, such as energy,
temperature, entropy, and other quantities depend highly on the gravity model.
Black hole thermodynamics is also closely related to quantum gravity, and
quantum effects of gravity might become very important in the region near the
horizon where gravity is strong. Therefore, the thermodynamics of the bumblebee
black holes might provide a new angle to probe Lorentz symmetry violation in the
strong field region.  In addition, the local thermodynamic stability of the
bumblebee black holes is an intriguing topic as well. It is usually
characterized by heat capacity which describes the ability of a thermodynamic
system to resist perturbations caused by exchanging a small amount of heat with
its environment.

Practically, the bumblebee theory can be interpreted as an extension of
Einstein-Maxwell theory by including a nonminimal coupling between the vector
field and gravity.  From this point of view, various solutions with vector hair,
such as spherical black holes, wormholes as well as solitons in asymptotically
flat or AdS spacetimes have been constructed as well \cite{Geng:2015kvs,
Heisenberg:2017hwb, Heisenberg:2017xda, Babichev:2017guv}. Moreover, the
thermodynamics of these vectorized black holes in various vector-tensor theories
have been well investigated using the Wald formalism \cite{Liu:2014tra,
Fan:2014ala, Fan:2016jnz, Fan:2017bka}. The Wald formalism, which was developed
by Wald \cite{Wald:1993nt, Iyer:1994ys}, gives a systematic procedure to
analytically derive the first law of thermodynamics for black holes in various
gravity theories. Especially, the first law of thermodynamics for the black
holes in the vector-tensor theory only containing the nonminimal coupling term
$B^\mu B^\mu R_{\mu\nu}$ has been analytically constructed by
\citet{Fan:2017bka}. By introducing the nonminimal coupling term, the first law
of thermodynamics is different from that of Reissner-Nordström (RN) black holes.
In addition to the Hawking temperature $T$ and the black hole entropy $S$, two
pairs of additional conjugated thermodynamic quantities were introduced, even
though such a vectorized black hole has only two 
free parameters. The thermodynamics and associated local stability of
bumblebee black holes are thus still interesting topics.

In this paper, based on the extended numerical black hole solutions constructed
by \citet{Xu:2022frb}, we investigate the associated thermodynamics numerically.
Especially, we construct a numerical version of the first law of thermodynamics
for the bumblebee black holes. In addition to the Hawking temperature and
entropy of the bumblebee black holes, we introduce only one pair of conjugated
thermodynamic quantities, denoted as $X$ and $Y$, where $Y$ can be interpreted
as an extension of the electric charge $Q$ of RN black holes. We further study
thermodynamic processes with $Y$ being constant and calculate the associated
heat capacity $C_Y$.

This paper is organized as follows. In Sec.~\ref{model}, we briefly overview the
construction of the spherical bumblebee black holes. In Sec.~\ref{firlaw}, after
calculating the Hawking temperature as well as the Wald entropy, we construct a
new version of the first law of thermodynamics for the bumblebee black holes. As
an example, we analyze thermodynamic processes where $Y$ is constant in the
entropy-charge parameter space. In Sec.~\ref{heatc}, we study the heat capacity
$C_Y$ as a potential indicator for thermodynamic stability of the bumblebee
black holes and further analyze deviations between bumblebee black holes and RN
black holes.  In Sec.~\ref{con}, we conclude this paper and give some further
comments. In this paper, we adopt the Planck-Gauss natural units; $G=c=\hbar =
k_{\rm B} = \epsilon_0 = 1$.

\section{Bumblebee black holes}\label{model}

Before investigating the thermodynamics of the spherical bumblebee black holes,
we first briefly overview how to construct them following the scheme given by
\citet{Xu:2022frb}. As we previously mentioned, we consider the vacuum bumblebee
theory with a nonzero bumblebee vector background $B^\mu = b^\mu$, which
minimizes the unspecified potential. With this in mind, by performing a
variation on the action in Eq.~\eqref{actionbum} with respect to the metric
$g^{\mu\nu}$ and the background bumblebee field $b_\mu$, we obtain the covariant
expressions for the equations of motion,
\begin{equation}\label{eom2}
R_{\mu\nu}-\frac{1}{2}g_{\mu\nu}R=\kappa T^{b}_{\mu\nu}\, , \quad D_\mu b^{\mu\nu}=\frac{\xi}{\kappa}R^{\mu\nu}b_\nu \, ,
\end{equation}
where the energy-momentum tensor contributed by the bumblebee field $b^\mu$ is \cite{Casana:2017jkc}
\begin{widetext}
\begin{eqnarray}
 T^{b}_{\mu\nu}  = && \frac{\xi}{2\ka} \bigg( g_{\mu\nu} b^\al b^\be R_{\al\be} - 2 b_\mu b_\la R_\nu^{\pt\nu \la}  -  2 b_\nu b_\la R_\mu^{\pt\mu \la}  + \Box_g ( b_\mu b_\nu )  - g_{\mn} D_\al D_\be ( b^\al b^\be )+  D_\ka D_\mu \left( b^\ka b_\nu \right) \cr
~\cr
&&  + D_\ka D_\nu ( b_\mu b^\ka )   + b_{\mu\la} b_\nu^{\pt\nu \la} - \frac{1}{4} g_\mn  b^{\al\be} b_{\al\be}  \bigg) + b_{\mu \lambda}b_\nu{}^{\lambda}-\frac{1}{4}g_{\mu\nu}b^{\mu\nu}b_{\mu\nu} \, , 
\end{eqnarray}
\end{widetext}
with $b_\mn = \prt_\mu b_\nu - \prt_\nu b_\mu$.  To construct a spherical black
hole, we consider the following spherical ansatz for $g_{\mu\nu}$ and $b_\mu$ in
$(t,r,\theta,\varphi)$ coordinates,
\begin{eqnarray}\label{sphe1}
&&\td s^2 = -h(r)\td t^2 + \frac{\td r^2}{f(r)}+ r^2 \left(\td \theta^2 + \sin^2 \theta \td \varphi^2 \right)  \, ,  \cr
~\cr
&&b_{\mu} = (b_t(r),0,0,0).
\end{eqnarray}
Here we consider the case where the bumblebee field only has the $t$ component
and it only depends on $r$ (see Ref.~\cite{Xu:2022frb} for extended solutions
with a nonzero radial component). Plugging the spherical ansatz in
Eq.~\eqref{sphe1} into the equations of motion, we have
\begin{widetext}
\begin{eqnarray}\label{eomfh}
0& = & -4h^2(h(-6+6f+rf')+r f (3 r \kappa b_t^2 +5f'))+2\xi r h (2 r f h b_t'^2 + b_t (b_t'(r h f' + f(4h+rh')) + 2 r f h +b_t'') \cr
&&-b_t^2((4 f + r f') h' + 2 r f h'')) \, , \cr
~\cr
0& = & 2h^2(-2 + 2 f + r f') - r^2 f h'^2 + r h ((6 f + r f')h' + 2 r f h'')  \, , \cr
~\cr
0&=&2 h(2(f-1)h^2 + r^2 \kappa b_t f b_t'h' + r h (f(r \kappa b_t'^2 +2 h')- \kappa b_t (b_t'(4 f + r f') + 2 r f b_t''))) \nn \\
&&+ r \xi b_t h(-2 r f b_t' h' + b_t ((4 f + r f')h' + 2 r f h'')) \, . 
\end{eqnarray}
\end{widetext}
Here the primes denote derivatives with respect to $r$. Denoting $r_h$ as the
location of the horizon, in the region near the horizon, we consider the
following asymptotic expansions for $h(r), \, f(r)$ and $b_t(r)$,
\begin{eqnarray}\label{ayho}
&& f(r) = f_1 (r - r_h) + f_2 (r -r_h)^2  + \cdots \, , \cr
~\cr
&& h(r)= h_1 (r - r_h) + h_2 (r -r_h)^2 + \cdots \, , \cr
~\cr
&& b_t(r)=b_{t0}+b_{t1}(r-r_h) + b_{t2}(r-r_h)^2 + \cdots .
\end{eqnarray}
Then the asymptotic expansions of Eq.~\eqref{eomfh}  give
\begin{widetext}
\begin{eqnarray}\label{nearh}
&&f_1 = \frac{4 h_1}{r_h (4 h_1 + b_{t1}^2 r_h (2\kappa -\xi ))} \, , \quad f_2 = \frac{-16h_1^2 \kappa +4 b_0^2 h_1 r_h \kappa (2 \kappa -\xi)-3b_{t1}^2 r_h^2 \xi (2\kappa + \xi)^2}{4 h_1 r_h^2 \kappa (4 h_1 +b_{t1}^2 r_h (2\kappa - \xi))} \, , \quad b_{t0} = 0 \, , \cr
&&~~  \\
&&h_2=-\frac{h_1}{r_h}+\frac{1}{4}b_{t1}^2(2\kappa - \xi) + \frac{b_{t1}^4 r_h \xi (2\kappa -\xi)^2}{16 h_1 \kappa} \, , \quad    b_{t2}=\frac{-16 b_{t1} h_1^2 \kappa + 2 b_{t1}^3 h_1 r_h (2\kappa - \xi)\xi + b_{t1}^5 r_h^2 \xi (-2\kappa + \xi)^2}{16 h_1^2 r_h \kappa} \, . \nn
\end{eqnarray}
\end{widetext}
Three coefficients $(h_1, b_{t1}, r_h)$ are not determined by the equations of
motion. Selecting appropriate values for these coefficients and performing
numerical integration using the equations of motion to a large enough $r$, we
obtain a class of numerical black hole solutions. In addition, for an
asymptotically flat bumblebee black hole, we expect that in the asymptotic
infinity, $h(r), \, f(r)$ and $b_t(r)$ have the following behavior,
\begin{eqnarray}\label{inf}
&&\left. h(r) \right|_{r \to \infty} = 1- \frac{2M}{r} + \frac{\tilde h_2}{r^2} + \cdots \, , \cr
~\cr
&&\left. f(r) \right|_{r \to \infty} = 1- \frac{2M}{r} + \frac{\tilde f_2}{r^2} + \cdots \, , \\
~\cr
&&\left. b_t(r) \right|_{r \to \infty} = \mu_\infty -\sqrt{\frac{2}{\kappa}}\frac{Q}{r} + \frac{\tilde b_2}{r^2} + \cdots  \, , \nn
\end{eqnarray}
where $\tilde h_2, \tilde f_2 $ and $\tilde b_2$ are determined by the equations
of motion. For given $(h_1, b_{t1}, r_h)$, one may find a black hole solution
with $\left.h(r)\right|_{r \to \infty} = \tilde h_0 \not = 1$. However, the
following rescaling transformation
\begin{equation}
h(r) \to \frac{h(r)}{\tilde h_0} \, , \quad b_t(r) \to \frac{b_t(r)}{\sqrt{\tilde h_0}} \, , \quad t \to \sqrt{\tilde h_0} \,  t\, ,
\end{equation}
keeps the equations of motion invariant, and can generate the desired black hole
solution with $\left.h(r)\right|_{r \to \infty} = 1$. Furthermore, it indicates
that the bumblebee black hole has only two parameters, the same as RN black
holes. From Eq.~\eqref{inf}, the mass $M$ and the vector charge $Q$ of the
bumblebee black hole read
\begin{equation}
M =- \frac{1}{2} \lim_{r \to \infty}  r^2 f'(r) \, , \quad Q = -\sqrt{\frac{\kappa}{2}} \lim_{r \to \infty} r^2 b_t' \, .
\end{equation}
Additionally, $\mu_\infty$ is the potential at infinity and it depends on $M$
and $Q$. When the coupling constant $\xi = 0$, namely that the bumblebee theory
reduces to the Einstein-Maxwell theory, the bumblebee black holes naturally
reduce to RN black holes with
\begin{eqnarray}
&& f(r)=h(r)=1-\frac{2M_{\rm RN}}{r} + \frac{Q_{\rm RN}^2}{r^2} \, ,
\nonumber \\
&& b_t(r)= \mu_{\rm RN} -\sqrt{\frac{2}{\kappa}}\frac{Q_{\rm RN}}{r}  \, .
\end{eqnarray}

Let us point out some essential differences between RN black holes and bumblebee
black holes. In the case of RN black holes, $\mu_{\rm RN} =
\sqrt{{2}/{\kappa}} \, {Q_{\rm RN}}/{r_h}$. In the Einstein-Maxwell theory,
there exists a $\text{U}(1)$ gauge symmetry. Therefore for RN black holes, one
can choose arbitrarily the zero point for the electric potential. Usually, it is
set at infinity so that $\mu_{\rm RN} = 0$.  The choice of $\mu_{\rm RN} =
\sqrt{{2}/{\kappa}} \, {Q_{\rm RN}}/{r_h}$ corresponds to setting the zero
point of the potential at the horizon. Furthermore, the electric charge $Q_{\rm
RN}$ is a conserved Noether charge associated with the $\text{U}(1)$ symmetry in
the Einstein-Maxwell theory. However, in the bumblebee theory, due to the
nonminimal coupling $\xi b^\mu b^\nu R_{\mu\nu}$, the $\text{U}(1)$ gauge
symmetry is broken.  Therefore the nonvanishing $\mu_\infty$ is not a gauge
choice any longer, but a source term for bumblebee black holes in addition to
the fact that the bumblebee vector charge $Q$ does not conserve when the
nonminimal coupling constant $\xi$ is nonzero.

Besides $\xi=0$, $\xi=2\ka$ is also a special value for the coupling constant because the solution is a Schwarzschild black hole with a nontrivial bumblebee vector. Such kind of Schwarzschild black holes with a nontrivial vector field has been found in other vector-tensor theories too
\cite{Bertolami:2005bh, Oliveira:2021abg, Gullu:2020qzu}. In the bumblebee
theory, they are simply
\begin{equation}\label{sch}
f(r) = h(r) = 1 - \frac{2M_{\rm Sch}}{r} \,, \quad b_t (r) = \mu_{\rm Sch}\left(1 - \frac{2M_{\rm Sch}}{r} \right) \, ,
\end{equation}
where $\mu_{\rm Sch}$ is an undetermined constant. In the 
Schwarzschild
black hole configuration, the bumblebee field can carry an arbitrarily large
vector charge, with the horizon of the black hole always being $r_h = 2M_{\rm
Sch}$.

\begin{figure}[t]
  \begin{center}
\includegraphics[width=0.5\textwidth, trim= 0 30 0 60,clip]{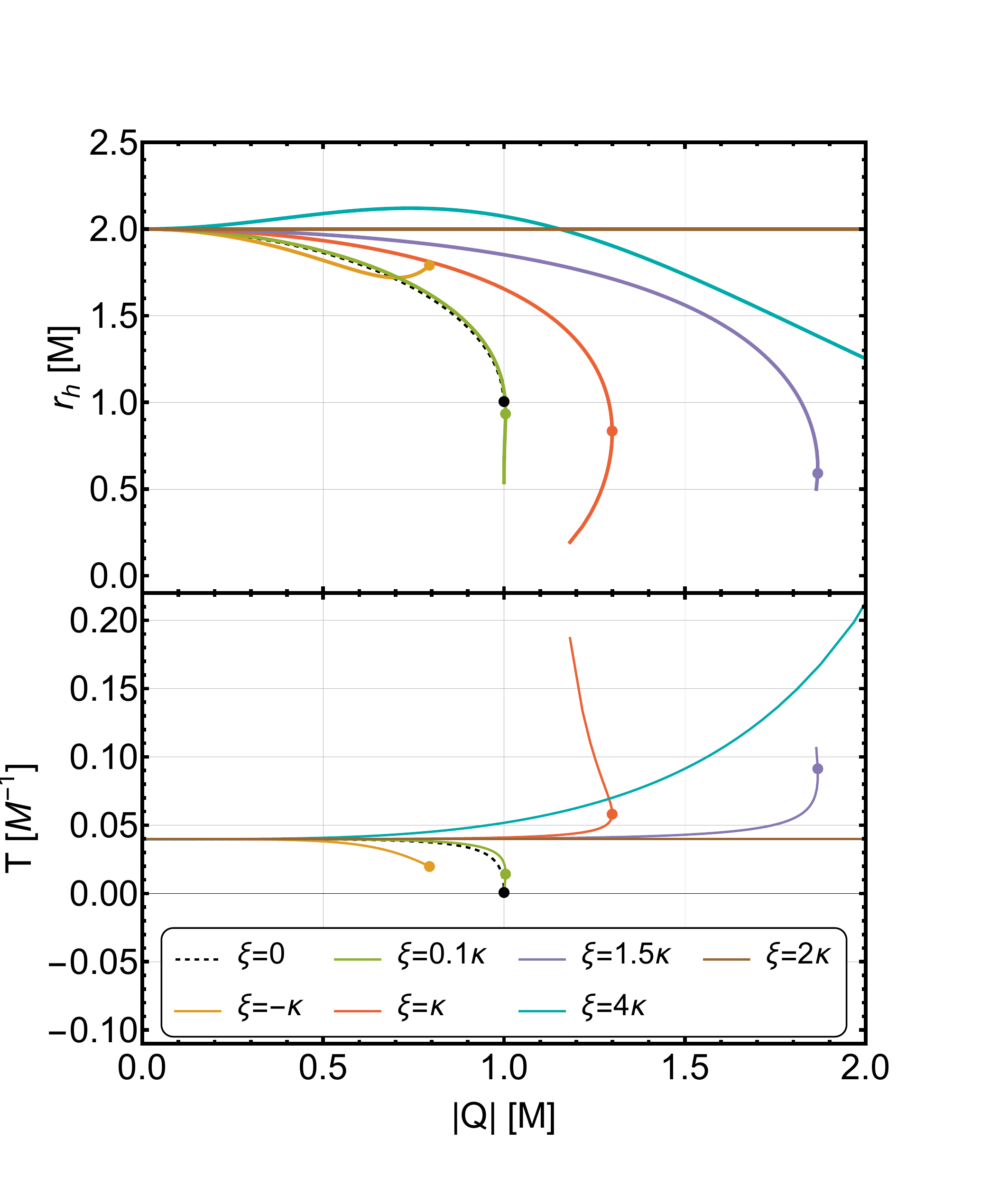}
  \caption{Upper panel: $r_h$ versus $Q$ for the bumblebee black holes. Lower
  panel: $T$ versus $Q$ for the bumblebee black holes. When $\xi<2\ka$,
  solutions with maximal charge to mass ratios are found and indicated by solid
  dots in the plots. } \label{rhQ}
  \end{center}
\end{figure}

For a general $\xi$, analytical black hole solutions are difficult to construct,
thus we investigate the bumblebee black holes numerically. In the upper panel of
Fig.~\ref{rhQ}, we show the relationship between the vector charge $Q$ and the
radius of the event horizon for bumblebee black holes. We present curves for seven values of $\xi$, namely $\xi = -\kappa, \, 0, \, 0.1\kappa,  \, \kappa, \, 1.5 \kappa, \, 2\kappa$, and $ 4\kappa$ as examples. For the case of $\xi = 0$, namely the RN
case, we have
\begin{equation}
\frac{r_{h_{\rm RN}}}{M_{\rm RN}} = 1+\sqrt{1-\frac{Q_{\rm RN}}{M_{\rm RN}}} \, ,
\end{equation}
which matches the black dashed line. For the cases of $0< \xi \lesssim 1.8
\kappa$, we find that the event horizon $r_h/M$ is a multivalued function with
respect to $Q/M$ in certain regions, namely that there exist two black hole solutions for the same
vector charge $Q$ and mass $M$.  It is clearly indicated by the case of $\xi =
\kappa$ shown as the red line in Fig.~\ref{rhQ}. Additionally, when $\xi <
2\kappa$, there exist maximal values for the vector charge of the bumblebee
black holes, namely maximal values for the charge to mass ratio. In the case of
RN black holes, the maximal charge to mass ratio is unity, corresponding to the
extremal RN black hole.  For the cases of $\xi < 0$, the maximal charge to mass
ratios of bumblebee black holes are less than unity, while for those of
$0<\xi<2\kappa$, the maximal charge to mass ratios are larger than unity. For
the case of $\xi=2\ka$, the horizon is at $r_h = 2 M$ regardless of the value of
the charge, so there is no maximal charge to mass ratio for those bumblebee
black holes. As for $\xi>2\ka$, we find that our numerical method can generate
solutions having very large charge to mass ratios but with unacceptable errors
when checking against the field equations. So it is still unclear if the
solutions end at finite charge to mass ratios or approach all the way to infinity.
%

\section{An extended first law of thermodynamics}\label{firlaw}

Black hole thermodynamics is an essential and rich topic in black hole physics,
connecting classical physics with quantum physics. In the Einstein-Maxwell
theory, an RN black hole obeys the following four laws of thermodynamics
\cite{Bardeen:1973gs}:
\begin{itemize}
\item \emph{The zeroth law}: The Hawking temperature $T$ is a constant on the
event horizon;
\item \emph{The first law}:  $\delta M_{\rm RN} = T_{\rm RN} \delta S_{\rm RN} +
\frac{Q_{\rm RN}}{\sqrt{4\pi}r_+} \delta Q_{\rm RN} $;
\item \emph{The second law}: The entropy of the black hole and its environment
does not decrease with time;
\item \emph{The third law}: The Hawking temperature $T$ can not be reduced to
zero by finite physical processes.
\end{itemize}
Taking RN black holes as an example, the Hawking temperature and the Bekenstein
entropy are \cite{Hawking:1975vcx},
\begin{equation}
T_{\rm RN} \equiv \frac{\kappa_{+_{\rm RN}}}{2\pi} \, , \quad S_{\rm RN} \equiv \frac{A_{\rm RN}}{4} \, .
\end{equation}
where $\kappa_{+_{\rm RN}} = ({r_+ - r_-})/{2 r_+^2}$ denotes the surface
gravity for RN black holes. Additionally, the area of an RN black hole is
$A_{\rm RN}=4\pi r_+^2$, where $r_\pm$ denotes the outer and inner horizon for
RN black holes,
\begin{equation}
r_\pm = M_{\rm RN} \pm \sqrt{M_{\rm RN}^2 -Q_{\rm RN}^2} \, .
\end{equation}
When $M_{\rm RN}=Q_{\rm RN}$, the Hawking temperature of a RN black hole
vanishes, and we called it an extremal black hole.

In the case of bumblebee black holes found by \citet{Xu:2022frb}, the zeroth law
is naturally satisfied since they are static and spherical. We then investigate
the first law of thermodynamics. Before achieving this, we first give the
Hawking temperature $T$ and the entropy $S$ of the bumblebee black holes. Since
the bumblebee theory involves a nonminimal interaction between the bumblebee
vector and the Ricci tensor, the entropy of the bumblebee black hole is
generalized according to Wald entropy as follows \cite{Wald:1993nt,
Iyer:1994ys},
\begin{equation}\label{wald}
S_W \equiv -2\pi \int_{r = r_h} \td^2 x \sqrt{h} ~ \frac{\partial {\cal L}}{\partial R^{\mu\nu\rho\sigma}} \epsilon^{\mu\nu} \epsilon^{\rho \sigma}\, ,
\end{equation}
where the integral is on the two-sphere of the horizon, $h$ is the determinant
of the metric of the two-sphere, and $\epsilon^{\mu\nu}$ is the antisymmetric
binormal to the horizon. Applying Eq.~\eqref{wald} to the spherical bumblebee
black holes, because the bumblebee field $b_t$ vanishes on the horizon according
to the expansion coefficients in Eq.~\eqref{nearh}, one find that $b_t$ has no
contribution to the Wald entropy $S_W$. Therefore the entropy of a bumblebee
black hole is still a quarter of its area, namely
\begin{equation}
S= \pi r_h^2 \, .
\end{equation}
Additionally, the Hawking temperature of the bumblebee black hole is 
\begin{equation}
T \equiv \frac{\kappa_+}{2\pi} = \frac{1}{2\sqrt{2} \, \pi} \left. \sqrt{(\nabla^\mu \eta^\nu)(\nabla_\mu \eta_\nu)} \right|_{r \to r_h} \, ,
\end{equation}
where $\eta^\mu \equiv \left(\partial/\partial t\right)^\mu$ represents the
timelike Killing vector. After applying the expansions for $h(r)$ and $f(r)$ in
Eq.~\eqref{ayho}, the Hawking temperature of the bumblebee black hole reads
\begin{equation}\label{tem}
T  = \frac{\sqrt{f_1 h_1}}{4\pi} \, .
\end{equation}

In the lower panel of Fig.~\ref{rhQ}, we show the relation between the
temperature $T$ and the vector charge $Q$ for the bumblebee black holes, for
several given values of $\xi$. The RN black holes, represented by the black
dashed line ($\xi = 0$), has the temperature decreasing to zero when the vector
charge approaches the maximal value, indicating the existence of the extremal RN
black hole.  For the other special case, namely the 
Schwarzschild black
holes with $\xi = 2\kappa$, the Hawking temperature is the same as that of the
Schwarzschild black hole and does not change with the vector charge.

For other values of $\xi$, we find that (i) the temperature never decreases to
zero as the vector charge grows if $\xi < \xi_c$ where $0.4 \kappa <\xi_c <
0.5\kappa$, (ii) the temperature does not decrease at all as the vector charge
increases if $\xi >\xi_c\,$. We mentioned that for $\xi<2\ka$, black holes with
maximal charge to mass ratios are found. But we see that these maximally charged
black holes do not have a vanishing temperature (except for the maximally
charged RN black hole), so we do not consider them as extremal black holes. The
fact that there are no extremal black holes defined by a zero temperature for
$\xi<2\ka$ $(\xi \ne 0)$ is unexpected, but might still be acceptable as the
temperature at least is finite. For $\xi>2\ka$, we find that the temperature (or
the factor $f_1$ precisely) goes to infinity when  
\begin{equation}
   | b_{t1} |\to b_c \equiv  \sqrt{\frac{4h_1}{r_h (\xi-2\kappa)}} \, ,
\end{equation}
as shown in Eq.~\eqref{nearh}. In fact, our numerical solutions seems to
indicate that as $|b_{t1}| \to b_c$, the mass and the charge of the black hole
go to infinity as well. But with the increasing numerical error as $|b_{t1}|$
approaches the critical value $b_c$, we cannot tell for sure, let alone the
limit for the charge to mass ratio as $|b_{t1}| \to b_c$.\footnote{This is why
we are unable to mark a maximally charged black hole solution in Fig.~\ref{rhQ}
for $\xi=4\ka$.}

With the temperature and the entropy found for the bumblebee black holes, we are
ready to construct the first law of thermodynamics for them.  A naive conjecture
is
\begin{equation} \label{fir1}
\delta M = T \delta S + \mu_\infty \delta Q \, ,
\end{equation}
which is a direct use of the first law for RN black holes. However,
Eq.~\eqref{fir1} turns out to be incorrect. As we previously mentioned, the
spherical bumblebee black holes have two free parameters, which can be chosen
from $(h_1,\, b_{t1},\, r_h)$, or more suitably here as the entropy $S$ and the
charge $Q$.  Then the mass of the black hole can be treated as a function of the
entropy $S$ and the vector charge $Q$, namely $M=M(S, Q)$, and 
\begin{equation}
\delta M = \left(\frac{\partial M}{\partial S}\right)_Q \delta S + \left(\frac{\partial M}{\partial Q}\right)_S \delta Q \, .
\end{equation}
In the case of $\xi = 0$ where the bumblebee black holes reduce to RN black
holes, we have 
\bea
&& M_{\rm RN} = \sqrt{\frac{S_{\rm RN}}{4\pi}} + \sqrt{\frac{\pi}{4S_{\rm RN}}} \, Q_{\rm RN}^2,
\nonumber \\
&& T_{\rm RN} = \frac{1}{4S_{\rm RN}} \left( \sqrt{\frac{S_{\rm RN}}{\pi}} - \sqrt{\frac{\pi}{S_{\rm RN}}} \, Q_{\rm RN}^2 \right) ,
\label{rnmt}
\eea
so that one can easily check
\begin{equation}
T_{\rm RN} = \left(\frac{\partial M_{\rm RN}}{\partial S_{\rm RN}}\right)_Q \, , \quad   \mu_{\rm RN} =\left(\frac{\partial M_{\rm RN}}{\partial Q_{\rm RN}}\right)_S \, .
\end{equation}
However, we numerically find that for a general $\xi \ne 0$,
\begin{equation}
T \not=  \left(\frac{\partial M}{\partial S}\right)_Q \, .
\end{equation}
Hence, Eq.~\eqref{fir1} does not hold in general.

An intuitive explanation for the failure of Eq.~\eqref{fir1} is that the
$\text{U}(1)$ symmetry is violated due to the nonminimal coupling term, so the
vector charge $Q$ is no longer naturally a good thermodynamic quantity to use.
Following the Wald formalism, the first law of thermodynamics for the bumblebee
black holes is actually \cite{Fan:2017bka}
\begin{eqnarray}\label{wald2}
\delta M &=& T \delta S + \mu_\infty \delta Q \cr
~\cr
&&- 2 \xi \left(\delta( Q\mu_\infty) -\frac{3}{2}M \delta (\mu_\infty)^2 - 4 \delta M \mu_\infty^2 \right) \, ,
\end{eqnarray}
where one has to introduce two pairs of conjugated thermodynamic quantities,
$(\mu_\infty, Q)$ and $(M,\mu_\infty^2)$. Of course, by treating $\mu_\infty$ as
a function of the two free parameters $S$ and $Q$, one is able to put
Eq.~\eqref{wald2} solely in terms of $\de S$ and $\de Q$ in principle, but the
coefficients before $\de S$ and $\de Q$ are no longer simply $T$ and
$\mu_\infty$.  

\begin{figure*}[t]
\includegraphics[width=0.8\textwidth]{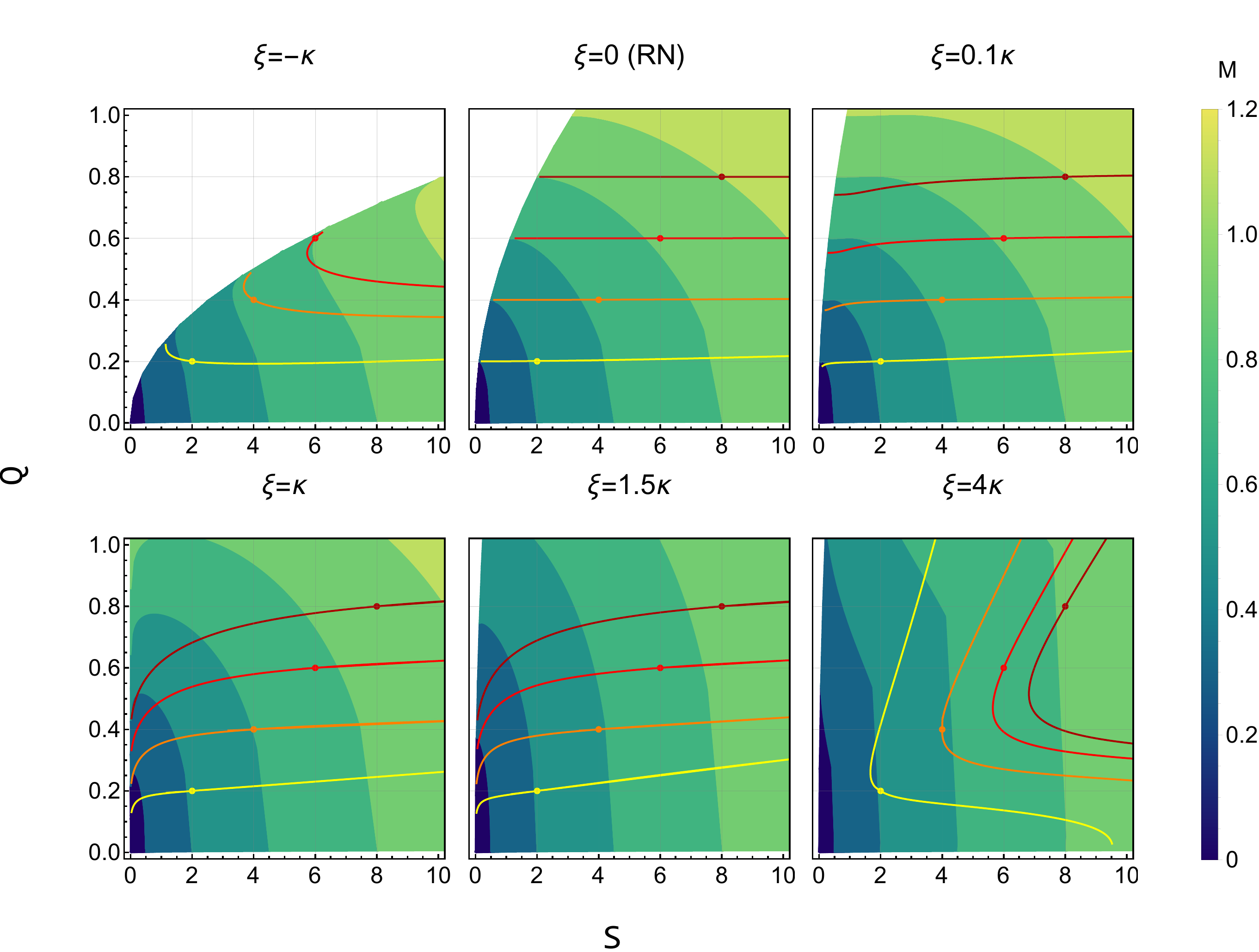}
\caption{Contours of the black hole mass $M$ as well as the family of curves of
constant-$Y$ processes on the $S$-$Q$ plane. Here we take $\xi = -\kappa, \, 0, \,
0.1\kappa, \, \kappa, \, 1.5\kappa$, and $4\kappa$ as examples. The yellow, orange, red and dark red lines correspond to constant-$Y$ curves passing through points $(S_0, Q_0) = (2,0.2), \, (4, 0.4), \, (6, 0.6)$ and $(8, 0.8)$ respectively.}\label{consY}
\end{figure*}

Our numerical solutions permit us to carry out the chain rule numerically to find out
$\de M$ in terms of $\de S$ and $\de Q$. But we prefer to find another quantity
$Y$, which is a function of $S$ and $Q$ so that the first law is in the familiar
form 
\begin{equation}\label{ther2}
\delta M = T \delta S + X \delta Y \, .
\end{equation}
In this sense, the quantities $X$ and $Y$ are the generalized vector potential
on the horizon and the generalized charge. For RN black holes, they reduce to
\begin{equation}
X = \frac{Q_{\rm RN}}{\sqrt{4\pi}r_+} \, , \quad Y = Q_{\rm RN} \, .
\end{equation}
In the following of this section, we describe how $X$ and
$Y$ might be calculated and then present our numerical results of contour plots for constant-$Y$
processes in the $S$-$Q$ parameter space. 

As $Y(S, Q)$ is a function of $S$ and $Q$, we have
\begin{equation}\label{ther1}
\de Y = X_1 \de S + X_2 \de Q \,,
\end{equation}
so that Eq.~\eqref{ther2} is 
\begin{equation}\label{dM}
\delta M = \left( T + XX_1 \right) \delta S + XX_2 \delta Q \, .
\end{equation}
The unknowns $X, \, X_1, \, X_2$ and $Y$, all treating as functions of $S$ and
$Q$, therefore satisfy
\begin{eqnarray}\label{pde}
 \left( \frac{\prt M}{\prt S} \right)_Q &=& T + X X_1 \, , \cr
~\cr
\left( \frac{\prt M}{\prt Q} \right)_S &=& X X_2 \, , \cr
~\cr
\frac{\prt^2 M}{\prt Q \prt S}  &=& \frac{\prt }{\prt Q} \bigg|_S \left( T + X X_1 \right) =  \frac{\prt }{\prt S} \bigg|_{Q} \left( XX_2 \right) , \cr
~\cr
\frac{\prt^2 Y}{\prt Q \prt S}  &=& \left( \frac{\prt X_1}{\prt Q} \right)_S =  \left( \frac{\prt X_2}{\prt S} \right)_Q .
\end{eqnarray}
Eliminating $X_1$ and $X_2$, we find a partial differential equation (PDE) for
$X(S, Q)$,
\begin{equation}\label{pdeforx}
\left( \frac{\prt M}{\prt Q} \right)_S X_S + \left( T-\left( \frac{\prt M}{\prt S} \right)_Q \right) X_Q = \left( \frac{\prt T}{\prt Q} \right)_S X \, ,
\end{equation}
where $X_S = \left(\prt X/\prt S\right)_Q$ and $X_Q = \left(\prt X/\prt
Q\right)_S$. Our numerical solutions give the functions $M(S, Q), \, T(S, Q)$ and their partial derivatives.
Together with a reasonable boundary condition, 
$X$ can be solved from Eq.~\eqref{pdeforx}. Once $X=X(S,Q)$ is known, we have
\bea
&& \left( \frac{\prt Y}{\prt S} \right)_Q = X_1 = \frac{1}{X}\left( \left( \frac{\prt M}{\prt S} \right)_Q - T \right), 
\nonumber\\
&& \left( \frac{\prt Y}{\prt Q} \right)_S = X_2 = \frac{1}{X} \left( \frac{\prt M}{\prt Q} \right)_S .
\eea
So $Y$ as a function of $S$ and $Q$ can be obtained by integrating Eq.~\rf{ther1} given an initial value $Y(S_0, Q_0)$.

Let us address that the quantities $X$ and $Y$ thus calculated are not unique. It is clear in Eq.~\rf{ther2} that if one pair of $X$ and $Y$ have been found, then $\tilde Y(Y)$ which is any reasonable function of $Y$ together with $\tilde X = X dY/d\tilde Y$ accomplishes the first law too, namely
\bea
\delta M = T \delta S + \tilde X \delta \tilde Y .
\eea
This freedom of transforming $Y$ to $\tilde Y(Y)$ is equivalent to the arbitrariness in specifying the boundary condition for $X$ in solving Eq.~\rf{pdeforx}, leading to numerous choices for $X$ and $Y$. Among all the choices, one might be able to fix $X$ and $Y$ up to a constant factor if the Smarr relation \cite{Smarr:1972kt}
\bea
M = 2TS + XY 
\label{smarr}
\eea
is required. However, the Smarr relation is not guaranteed to be consistent with Eq.~\rf{pdeforx}. It turns out to require the following condition (see Appendix~\ref{app1} for the derivation)
\bea
0 &=& 2S \left( \left( \frac{\prt M}{\prt Q} \right)_S\left( \frac{\prt T}{\prt S} \right)_Q - \left( \frac{\prt M}{\prt S} \right)_Q\left( \frac{\prt T}{\prt Q} \right)_S \right)
\nonumber \\
&& + T \left( \frac{\prt M}{\prt Q} \right)_S + M \left( \frac{\prt T}{\prt Q} \right)_S   .
\label{smarrcon}
\eea
One can check that RN black holes satisfy the condition using the analytical relation in Eq.~\rf{rnmt}. But for numerical bumblebee black hole solutions, we cannot say for sure if the condition is satisfied as small residuals are found in our numerical calculation.

Due to the lack of a unique choice for $X$ and $Y$, we quit calculating them but focus on investigating processes with $Y$ (thus also any $\tilde Y(Y)$) being constant. These processes are curves on the $S$-$Q$ plane defined by
\begin{equation}\label{caly}
\left( \frac{\td Q}{\td S} \right)_Y = -\frac{X_1}{X_2} = \frac{ T - \left(\frac{\partial M}{\partial S}\right)_Q }{ \left(\frac{\partial M}{\partial Q} \right)_{S} } \, .
\end{equation}
In Fig.~\ref{consY}, the colored regions show contours of the black hole mass
$M$, while the curves are solutions to Eq.~\rf{caly}, namely isolines of the generalized charge $Y$ in the $S$-$Q$
plane. Note that black hole solutions do not exist in the blank zones in the
plots.  To numerically obtain the curves representing the constant-$Y$ processes
in Fig.~\ref{consY}, we start with an initial point $(S_0, Q_0)$ on the plane
and proceed to the next point by the following recursive relation 
\begin{equation}
Q_{i+1}=Q_i + \frac{T_i - \left(\frac{\bar{M}_{i}- M_{i}}{\bar S_{i}- S_i}\right)_{Q_i}}{\left(\frac{\tilde M_{i}- M_i}{\tilde Q_{i}- Q_i}\right)_{S_{i}}} (S_{i+1}-S_i)  \, .
\end{equation}
Here $\bar M_{i}$ and $\bar S_{i}$ denote the mass and entropy of black holes
sharing the same vector charge $Q_i$ with those of mass $M_{i}$ and entropy
$S_i$. Similarly, $\tilde M_{i}$ and $\tilde Q_{i}$ denote the mass and vector
charge of those black holes sharing the same entropy $S_i$ with those of mass
$M_i$ and charge $Q_i$. To ensure numerical accuracy, we choose the step
$S_{i+1} - S_i \approx 10^{-2}$ (in the Planck-Gauss natural units). Following
this scheme step-by-step, we obtain a series of points revealing a constant-$Y$
process on the $S$-$Q$ plane. Varying the starting point, we can obtain a family of
constant-$Y$ processes for different values of $Y$.  As shown in
Fig.~\ref{consY}, we choose $(S_0, Q_0)=(2,0.2), (4,0.4), (6,0.6), (8,0.8)$ as
four initial points, and calculated four typical constant-$Y$ curves in the
$S$-$Q$ plane for several values of $\xi$. 

We notice that in the case of $\xi = 0$ (RN black hole), the curves of
constant-$Y$ processes are horizontal straight lines, consistent with the fact
$Y = Q_{\rm RN}$ for RN black holes.  For other values of $\xi$, we
find that the lines of constant-$Y$ processes are not parallel anymore. The
larger $|\xi|$ is, the more curved the lines are. For example, when $\xi =4
\kappa$, the constant-$Y$ lines are so bent that for a given entropy $S$, there
are generally two values of $Q$ corresponding to the same value of $Y$, as shown
in the third plot in the bottom panel of Fig.~\ref{consY}.

\section{local thermodynamic stability}\label{heatc}

After revealing constant-$Y$ processes in the $S$-$Q$ plane, now we focus on the
local thermodynamic stability which is indicated by the heat capacity of a
thermodynamic system in its thermodynamic equilibrium \cite{Avramov:2023eif}. 
In general, heat capacity is defined as the heat transferred per unit change of
temperature, 
\begin{equation}
    C \equiv \frac{\delta {\cal Q} }{\delta T} \, .
\end{equation}
Heat capacity depends on the specific thermodynamic process under consideration.
Different thermodynamic processes give different heat capacities. For example,
two commonly used heat capacities for an ideal gas system are under constant
volume (isochoric) and under constant pressure (isobaric) correspondingly. A
positive heat capacity, $C>0$, indicates that the system in equilibrium is
locally thermodynamically stable, while a negative heat capacity, $C<0$,
indicates that it is locally thermodynamically unstable. In general, most
matters, such as ideal gas, liquid, and solid have positive heat capacities. 
Supposing that a thermodynamic system composed of ideal gas is in thermodynamic
equilibrium with its environment, when the temperature of the environment
raises, the system of ideal gas will absorb heat from the environment.  Since
the heat capacity of ideal gas is positive, through absorbing heat from the
environment, the temperature of the ideal gas will increase and finally reach a
new equilibrium with the environment. Thus the ideal gas system is
locally thermodynamically stable. 

However, self-gravitating systems are different. In black hole thermodynamics,
black holes can have negative heat capacities, implying that they are locally
thermodynamically unstable. For example, the heat capacity of the Schwarzschild
black hole is 
\begin{equation}\label{hcsch}
    C_{\rm  Sch} = \frac{\delta {\cal Q}}{\delta T} = \frac{\td M}{\td T} = -8\pi M^2 \, .
\end{equation}
The negative heat capacity means that a Schwarzschild black hole is locally
thermodynamically unstable, reflecting the fact that a Schwarzschild black hole
will continuously release energy through Hawking radiation, the so-called black
hole evaporation \cite{Traschen:1999zr}. As for the RN black holes, considering
constant-$Q$ processes which are analogs of constant-volume processes for ideal
gas, the associated heat capacity is \cite{Altamirano:2014tva}
\begin{eqnarray} \label{crn}
C_Q &=& \left(\frac{\td M}{\td T} \right)_{Q}\cr
&& ~ \\
&=& -\frac{2 M^2 \pi (1+\sqrt{1-\lambda^2})^2 (1-\lambda^2 + \sqrt{1-\lambda^2})}{1-2\lambda^2 + \sqrt{1-\lambda^2}} \, , \nn
\end{eqnarray}
where $\lambda \equiv |Q_{\rm RN}|/M_{\rm RN}$. It is easy to observe that there
is a discontinuous point when $1-2\lambda^2 + \sqrt{1-\lambda^2} =0$, giving
$\lambda =\sqrt{3}/2$. When $\lambda < \sqrt{3}/2$, the heat capacity of an RN
black hole is negative, giving a locally thermodynamically unstable charged
black hole, while when $\lambda > \sqrt{3}/2$, the heat capacity turns to be
positive through the discontinuous point, giving a locally thermodynamically
stable charged black hole.  Additionally, such a change of the sign for heat
capacity due to a divergence indicates a discontinuous phase transition
\cite{Avramov:2023eif}. 
\begin{figure}[t]
  \begin{center}
\includegraphics[width=0.45\textwidth, trim= 0 45 0 45,clip]{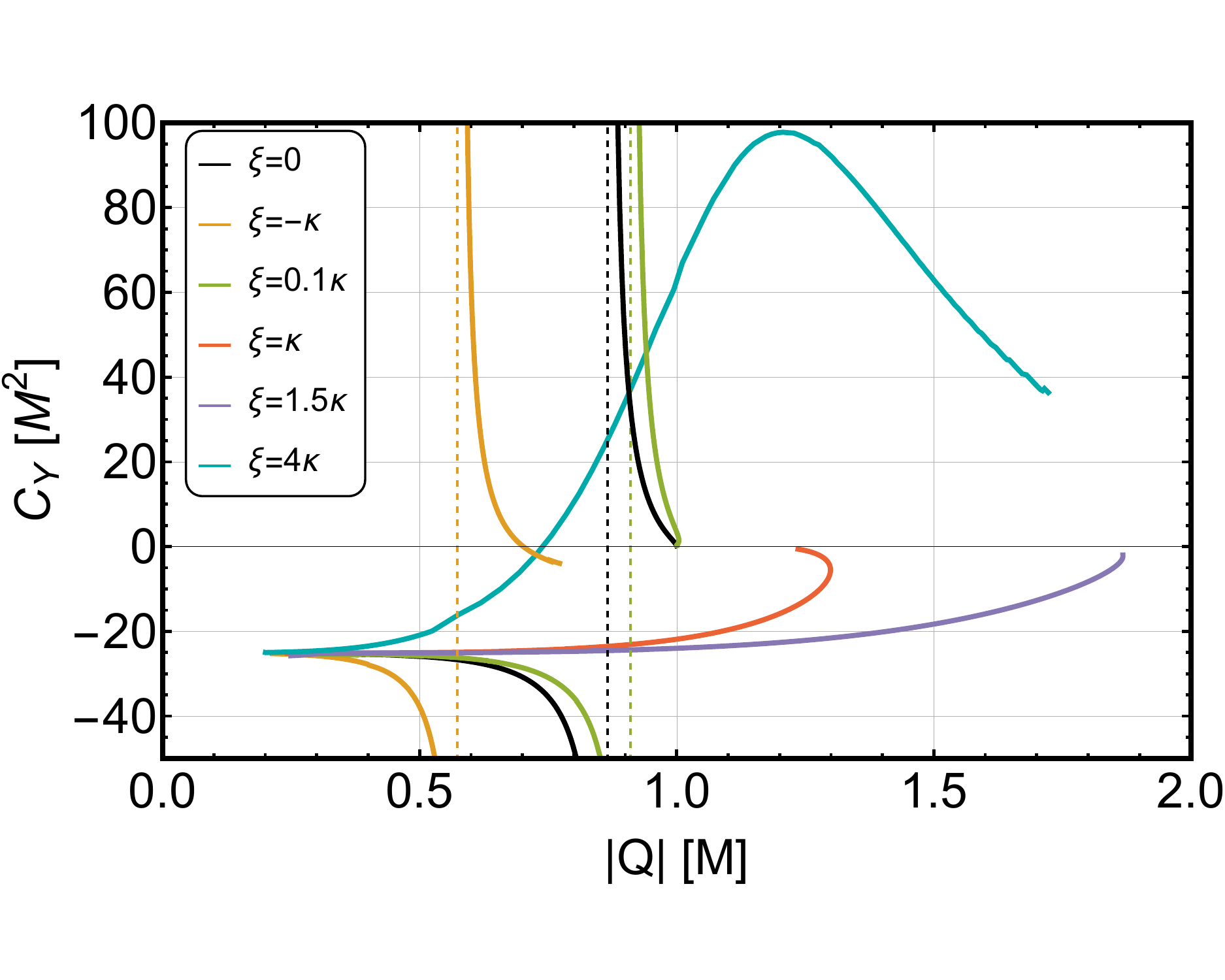}
  \caption{Relation between the heat capacity associated with constant-$Y$
  processes and the charge for the bumblebee black holes.} \label{cysum}
  \end{center}
\end{figure}

Based on the constant-$Y$ processes we discussed in the previous section,
Eq.~\eqref{crn} can be extended for the bumblebee black holes as $Y$ is the
generalized charge appearing in the first law of thermodynamics. The heat
capacity for constant-$Y$ processes can be calculated as 
\begin{equation}
    C_Y = \left(\frac{\td M}{\td T} \right)_{Y} \, .
\end{equation}
Since $C_Y$ has no known analytic expression, we numerically calculate $C_Y$ for
various $\xi$ and present the results as functions of the charge in
Fig.~\ref{cysum}. 

First, Fig.~\ref{cysum} indicates that for each $\xi$, when $|Q| \to 0$, the
heat capacity of the bumblebee black holes becomes $C_Y/M^2 \to -8 \pi \approx
-25.12$, which is consistent with the fact that the bumblebee black holes reduce
to the Schwarzschild black holes when the vector charge vanishes.  We are not
able to show in Fig.~\ref{cysum} the value of $C_Y/M^2$ at $Q=0$ as the
numerical error prevents our calculation when $|Q|/M$ is very small. But the
trend for $C_Y/M^2 \to -8 \pi \approx -25.12$ is clear.  

Second, when $\xi <0$ (taking $\xi=-\ka$ as an example), we find that similar to RN black holes, there exist a
divergent point where the sign of the heat capacity changes. A closer inspection finds that the
charge to mass ratio at the discontinuity is smaller than that of RN black holes
($Q/M \approx 0.52 $ for $\xi=-\ka$).  In addition to the discontinuous phase
transition, when $Q/M$ is large enough, we find that there exists a continuous
phase transition point where the heat capacity changes the sign smoothly.
Therefore, for $\xi < 0$, a bumblebee black hole begins as a locally
thermodynamically unstable black hole, turns to stable through a discontinuous
phase transition, and then turns to unstable again through a smooth phase
transition as the vector charge grows. 

Third, when $\xi>0$, we find that the divergence of the heat capacity, if
exists, appears on the right of the divergence point of the RN black holes. As
$\xi$ increases, we find that the divergence point disappears at $\xi \to
\xi_c$, where $0.4\ka < \xi_c < 0.5\ka$ is the same critical value found in Sec.~\ref{firlaw}.
When $ \xi_c <\xi < 2 \kappa$ (for example, $\xi = \kappa$ and $1.5\kappa$ in Fig.~\ref{cysum}), all bumblebee black holes are locally
thermodynamically unstable and do not have any phase transition point. 

As for the case of $\xi=2\kappa$ where the bumblebee black hole reduces to the
Schwarzschild black hole, Eq.~\eqref{sch} the generalized charge $Y$ is undefined.
There is only one definition of heat capacity, which is Eq.~\eqref{hcsch}.

Finally, when $\xi >2\kappa$ (taking $\xi = 4\kappa$ as an example), we find
that there exists only a continuous phase transition point ($Q/M \approx 0.82$
for $\xi = 4\kappa$). The bumblebee black hole is locally thermodynamically
unstable when the charge to mass ratio is small and is stable when the charge to
mass ratio is large.

\section{conclusion and discussion}\label{con}

The bumblebee theory is not only one of the modified gravity theories with
Lorentz symmetry violation but also a potential extension of the
Einstein-Maxwell theory. Similar to RN black holes, a spherical black hole with
a nontrivial bumblebee field can also be described by its mass $M$ and charge
$Q$. We have numerically constructed the first law of thermodynamics for the
bumblebee black holes by introducing a pair of new thermodynamic quantities $X$
and $Y$, which can be interpreted as generalizations of the electric potential
$\mu_{\rm RN}$ and charge $Q_{\rm RN}$ of RN black holes respectively.
Algorithms to calculate $X$ and $Y$ in terms of the entropy $S$ and the charge
$Q$ have been presented and constant-$Y$ processes in the $S$-$Q$ plane have
been constructed.

To investigate the local thermodynamic stability of the bumblebee black holes,
we have further calculated the heat capacity associated with the constant-$Y$
processes. When $0<\xi \lesssim 0.4\kappa $, we find that there exists a
divergence with the changing of the sign for the heat capacity, indicating a
discontinuous phase transition from locally thermodynamically unstable black
holes to stable black holes. When $\xi < 0$, in addition to the discontinuous
phase transition, we also find that when the vector charge is sufficiently
large, there exists of a continuous phase transition where the bumblebee black
hole becomes locally thermodynamically stable again. As for the case of $
0.4\kappa  \lesssim \xi < 2 \kappa $, the divergence of heat capacity disappears
and the bumblebee black holes are locally thermodynamically unstable for all
charge to mass ratios. When $\xi > 2\kappa$, there is no divergence but a smooth
phase transition point that separates locally thermodynamically stable and
unstable bumblebee black holes. 

In modified gravity, the Wald formalism is usually used to analytically investigate black hole thermodynamics. However, the first law of thermodynamics given by the Wald
formalism needs to introduce additional conjugated thermodynamic quantities,
usually more than the number of the free parameters in the black hole solutions.
Therefore, we try another path that numerically constructs the first law of
thermodynamics, using the bumblebee theory as an example. Its correspondence to
the Wald formalism still needs further investigation. 

In addition, we remark that we here focus on the local thermodynamic properties
of the bumblebee black holes, but the global stability is still unclear. The
global and local thermodynamic properties are not equivalent in black hole
thermodynamics since black holes generally have negative heat capacity.  In
principle, recall that $\left. b_t \right|_{r \to \infty} \not= 0 $ in
Eq.~\eqref{inf}, and the bumblebee vector field can be treated as sourced at
infinity, namely that our bumblebee black holes are globally grand canonical
ensembles.  To investigate the global thermodynamic stability of the bumblebee
black holes, one needs to calculate the free energy by fixing $T$ and
$\mu_\infty$ \cite{Mai:2020sac}. 
Furthermore, as also mentioned in Ref.~\cite{Mai:2020sac}, black holes in astrophysical environments, usually with accretion disk and interaction with other matters, are globally grand canonical ensembles. The global thermodynamic stability should be considered. The astrophysical properties of the bumblebee black holes have been studied in Ref.~\cite{Xu:2023xqh}. In this paper, we mainly focus on the theoretical properties of the bumblebee black hole thermodynamics. The application of the thermodynamics studied in the current work to astrophysical black holes is unclear at the current stage. We thus leave the application aspect of our work to future studies.

\begin{acknowledgments}
We thank Run-Qiu Yang and Hong L\"{u} for useful discussions.  This work
was supported by the National Natural Science Foundation of China (Grants~No.~12247128,
N0.~11991053, No.~11975027), the China Postdoctoral Science Foundation (No.~2021TQ0018), the
National SKA Program of China (No.~2020SKA0120300), the Max Planck Partner Group
Program funded by the Max Planck Society, and the High-Performance Computing
Platform of Peking University.  R. X. is supported by the Boya postdoctoral
fellowship at Peking University.
\end{acknowledgments}

\appendix

\section{A necessary condition for the Smarr relation}
\label{app1}

Here we derive Eq.~\rf{smarrcon} under the assumption of the Smarr relation in Eq.~\rf{smarr} with $X$ and $Y$ satisfying the first law in Eq.~\rf{ther2}. 
Using the Smarr relation, the derivative of $X$ with respect to $S$ along the constant-$Y$ processes defined by Eq.~\rf{caly} is
\bea
Y \left( \frac{\td X}{\td S} \right)_Y  &=& \left( \frac{\td}{\td S} \left( M-2TS \right) \right)_Y 
\nonumber \\
&=&  \left( \frac{\prt M}{\prt S} \right)_Q - 2S \left( \frac{\prt T}{\prt S} \right)_Q - 2T 
\nonumber \\
&& + \left( \left( \frac{\prt M}{\prt Q} \right)_S - 2S \left( \frac{\prt T}{\prt Q} \right)_S \right) \left( \frac{\td Q}{\td S} \right)_Y \, .
\label{dxds1}
\eea
In the meanwhile, $X$ satisfies Eq.~\rf{pdeforx}, giving 
\bea
\left( \frac{\td X}{\td S} \right)_Y &=& \left( \frac{\prt X}{\prt S} \right)_Q + \left( \frac{\prt X}{\prt Q} \right)_S \left( \frac{\td Q}{\td S} \right)_Y 
\nonumber \\
&=& \frac{X \left( \frac{\prt T}{\prt Q} \right)_S}{\left( \frac{\prt M}{\prt Q} \right)_S} \,,
\label{dxds2}
\eea
where Eq.~\rf{caly} has been used. For Eq.~\rf{dxds1} and Eq.~\rf{dxds2} to be consistent with each other, we find 
\bea
&& \left( \frac{\prt M}{\prt Q} \right)_S \left( \left( \frac{\prt M}{\prt S} \right)_Q - 2S \left( \frac{\prt T}{\prt S} \right)_Q - 2T \right) 
\nonumber \\
&& + \left( \left( \frac{\prt M}{\prt Q} \right)_S - 2S \left( \frac{\prt T}{\prt Q} \right)_S \right) \left( T - \left( \frac{\prt M}{\prt S} \right)_Q \right)
\nonumber \\
&=& \left( M-2TS\right) \left( \frac{\prt T}{\prt Q} \right)_S \,,
\eea
which simplifies to Eq.~\rf{smarrcon}.

\bibliography{bum_thermo}

\begin{thebibliography}{33}%
\makeatletter
\providecommand \@ifxundefined [1]{%
 \@ifx{#1\undefined}
}%
\providecommand \@ifnum [1]{%
 \ifnum #1\expandafter \@firstoftwo
 \else \expandafter \@secondoftwo
 \fi
}%
\providecommand \@ifx [1]{%
 \ifx #1\expandafter \@firstoftwo
 \else \expandafter \@secondoftwo
 \fi
}%
\providecommand \natexlab [1]{#1}%
\providecommand \enquote  [1]{``#1''}%
\providecommand \bibnamefont  [1]{#1}%
\providecommand \bibfnamefont [1]{#1}%
\providecommand \citenamefont [1]{#1}%
\providecommand \href@noop [0]{\@secondoftwo}%
\providecommand \href [0]{\begingroup \@sanitize@url \@href}%
\providecommand \@href[1]{\@@startlink{#1}\@@href}%
\providecommand \@@href[1]{\endgroup#1\@@endlink}%
\providecommand \@sanitize@url [0]{\catcode `\\12\catcode `\$12\catcode
  `\&12\catcode `\#12\catcode `\^12\catcode `\_12\catcode `\%12\relax}%
\providecommand \@@startlink[1]{}%
\providecommand \@@endlink[0]{}%
\providecommand \url  [0]{\begingroup\@sanitize@url \@url }%
\providecommand \@url [1]{\endgroup\@href {#1}{\urlprefix }}%
\providecommand \urlprefix  [0]{URL }%
\providecommand \Eprint [0]{\href }%
\providecommand \doibase [0]{http://dx.doi.org/}%
\providecommand \selectlanguage [0]{\@gobble}%
\providecommand \bibinfo  [0]{\@secondoftwo}%
\providecommand \bibfield  [0]{\@secondoftwo}%
\providecommand \translation [1]{[#1]}%
\providecommand \BibitemOpen [0]{}%
\providecommand \bibitemStop [0]{}%
\providecommand \bibitemNoStop [0]{.\EOS\space}%
\providecommand \EOS [0]{\spacefactor3000\relax}%
\providecommand \BibitemShut  [1]{\csname bibitem#1\endcsname}%
\let\auto@bib@innerbib\@empty
\bibitem [{\citenamefont {Kosteleck\'y}(2004)}]{Kostelecky:2003fs}%
  \BibitemOpen
  \bibfield  {author} {\bibinfo {author} {\bibfnamefont {V.~A.}\ \bibnamefont
  {Kosteleck\'y}},\ }\href {\doibase 10.1103/PhysRevD.69.105009} {\bibfield
  {journal} {\bibinfo  {journal} {Phys. Rev. D}\ }\textbf {\bibinfo {volume}
  {69}},\ \bibinfo {pages} {105009} (\bibinfo {year} {2004})},\ \Eprint
  {http://arxiv.org/abs/hep-th/0312310} {arXiv:hep-th/0312310} \BibitemShut
  {NoStop}%
\bibitem [{\citenamefont {Kosteleck\'y}\ and\ \citenamefont
  {Mewes}(2009)}]{Kostelecky:2009zp}%
  \BibitemOpen
  \bibfield  {author} {\bibinfo {author} {\bibfnamefont {V.~A.}\ \bibnamefont
  {Kosteleck\'y}}\ and\ \bibinfo {author} {\bibfnamefont {M.}~\bibnamefont
  {Mewes}},\ }\href {\doibase 10.1103/PhysRevD.80.015020} {\bibfield  {journal}
  {\bibinfo  {journal} {Phys. Rev. D}\ }\textbf {\bibinfo {volume} {80}},\
  \bibinfo {pages} {015020} (\bibinfo {year} {2009})},\ \Eprint
  {http://arxiv.org/abs/0905.0031} {arXiv:0905.0031 [hep-ph]} \BibitemShut
  {NoStop}%
\bibitem [{\citenamefont {Kosteleck\'y}\ and\ \citenamefont
  {Mewes}(2012)}]{Kostelecky:2011gq}%
  \BibitemOpen
  \bibfield  {author} {\bibinfo {author} {\bibfnamefont {A.}~\bibnamefont
  {Kosteleck\'y}}\ and\ \bibinfo {author} {\bibfnamefont {M.}~\bibnamefont
  {Mewes}},\ }\href {\doibase 10.1103/PhysRevD.85.096005} {\bibfield  {journal}
  {\bibinfo  {journal} {Phys. Rev. D}\ }\textbf {\bibinfo {volume} {85}},\
  \bibinfo {pages} {096005} (\bibinfo {year} {2012})},\ \Eprint
  {http://arxiv.org/abs/1112.6395} {arXiv:1112.6395 [hep-ph]} \BibitemShut
  {NoStop}%
\bibitem [{\citenamefont {Bertolami}\ and\ \citenamefont
  {Paramos}(2005)}]{Bertolami:2005bh}%
  \BibitemOpen
  \bibfield  {author} {\bibinfo {author} {\bibfnamefont {O.}~\bibnamefont
  {Bertolami}}\ and\ \bibinfo {author} {\bibfnamefont {J.}~\bibnamefont
  {Paramos}},\ }\href {\doibase 10.1103/PhysRevD.72.044001} {\bibfield
  {journal} {\bibinfo  {journal} {Phys. Rev. D}\ }\textbf {\bibinfo {volume}
  {72}},\ \bibinfo {pages} {044001} (\bibinfo {year} {2005})},\ \Eprint
  {http://arxiv.org/abs/hep-th/0504215} {arXiv:hep-th/0504215} \BibitemShut
  {NoStop}%
\bibitem [{\citenamefont {Oliveira}\ \emph {et~al.}(2021)\citenamefont
  {Oliveira}, \citenamefont {Dantas},\ and\ \citenamefont
  {Almeida}}]{Oliveira:2021abg}%
  \BibitemOpen
  \bibfield  {author} {\bibinfo {author} {\bibfnamefont {R.}~\bibnamefont
  {Oliveira}}, \bibinfo {author} {\bibfnamefont {D.~M.}\ \bibnamefont
  {Dantas}}, \ and\ \bibinfo {author} {\bibfnamefont {C.~A.~S.}\ \bibnamefont
  {Almeida}},\ }\href {\doibase 10.1209/0295-5075/ac130c} {\bibfield  {journal}
  {\bibinfo  {journal} {EPL}\ }\textbf {\bibinfo {volume} {135}},\ \bibinfo
  {pages} {10003} (\bibinfo {year} {2021})},\ \Eprint
  {http://arxiv.org/abs/2105.07956} {arXiv:2105.07956 [gr-qc]} \BibitemShut
  {NoStop}%
\bibitem [{\citenamefont {G\"ull\"u}\ and\ \citenamefont
  {\"Ovg\"un}(2022)}]{Gullu:2020qzu}%
  \BibitemOpen
  \bibfield  {author} {\bibinfo {author} {\bibfnamefont {I.}~\bibnamefont
  {G\"ull\"u}}\ and\ \bibinfo {author} {\bibfnamefont {A.}~\bibnamefont
  {\"Ovg\"un}},\ }\href {\doibase 10.1016/j.aop.2021.168721} {\bibfield
  {journal} {\bibinfo  {journal} {Annals Phys.}\ }\textbf {\bibinfo {volume}
  {436}},\ \bibinfo {pages} {168721} (\bibinfo {year} {2022})},\ \Eprint
  {http://arxiv.org/abs/2012.02611} {arXiv:2012.02611 [gr-qc]} \BibitemShut
  {NoStop}%
\bibitem [{\citenamefont {Maluf}\ and\ \citenamefont
  {Neves}(2021)}]{Maluf:2020kgf}%
  \BibitemOpen
  \bibfield  {author} {\bibinfo {author} {\bibfnamefont {R.~V.}\ \bibnamefont
  {Maluf}}\ and\ \bibinfo {author} {\bibfnamefont {J.~C.~S.}\ \bibnamefont
  {Neves}},\ }\href {\doibase 10.1103/PhysRevD.103.044002} {\bibfield
  {journal} {\bibinfo  {journal} {Phys. Rev. D}\ }\textbf {\bibinfo {volume}
  {103}},\ \bibinfo {pages} {044002} (\bibinfo {year} {2021})},\ \Eprint
  {http://arxiv.org/abs/2011.12841} {arXiv:2011.12841 [gr-qc]} \BibitemShut
  {NoStop}%
\bibitem [{\citenamefont {Izmailov}\ and\ \citenamefont
  {Nandi}(2022)}]{Izmailov:2022jon}%
  \BibitemOpen
  \bibfield  {author} {\bibinfo {author} {\bibfnamefont {R.~N.}\ \bibnamefont
  {Izmailov}}\ and\ \bibinfo {author} {\bibfnamefont {K.~K.}\ \bibnamefont
  {Nandi}},\ }\href {\doibase 10.1088/1361-6382/ac8fda} {\bibfield  {journal}
  {\bibinfo  {journal} {Class. Quant. Grav.}\ }\textbf {\bibinfo {volume}
  {39}},\ \bibinfo {pages} {215006} (\bibinfo {year} {2022})}\BibitemShut
  {NoStop}%
\bibitem [{\citenamefont {Liu}\ \emph {et~al.}(2023)\citenamefont {Liu},
  \citenamefont {Fang}, \citenamefont {Jing},\ and\ \citenamefont
  {Wang}}]{Liu:2022dcn}%
  \BibitemOpen
  \bibfield  {author} {\bibinfo {author} {\bibfnamefont {W.}~\bibnamefont
  {Liu}}, \bibinfo {author} {\bibfnamefont {X.}~\bibnamefont {Fang}}, \bibinfo
  {author} {\bibfnamefont {J.}~\bibnamefont {Jing}}, \ and\ \bibinfo {author}
  {\bibfnamefont {J.}~\bibnamefont {Wang}},\ }\href {\doibase
  10.1140/epjc/s10052-023-11231-5} {\bibfield  {journal} {\bibinfo  {journal}
  {Eur. Phys. J. C}\ }\textbf {\bibinfo {volume} {83}},\ \bibinfo {pages} {83}
  (\bibinfo {year} {2023})},\ \Eprint {http://arxiv.org/abs/2211.03156}
  {arXiv:2211.03156 [gr-qc]} \BibitemShut {NoStop}%
\bibitem [{\citenamefont {Xu}\ \emph {et~al.}(2023{\natexlab{a}})\citenamefont
  {Xu}, \citenamefont {Liang},\ and\ \citenamefont {Shao}}]{Xu:2022frb}%
  \BibitemOpen
  \bibfield  {author} {\bibinfo {author} {\bibfnamefont {R.}~\bibnamefont
  {Xu}}, \bibinfo {author} {\bibfnamefont {D.}~\bibnamefont {Liang}}, \ and\
  \bibinfo {author} {\bibfnamefont {L.}~\bibnamefont {Shao}},\ }\href {\doibase
  10.1103/PhysRevD.107.024011} {\bibfield  {journal} {\bibinfo  {journal}
  {Phys. Rev. D}\ }\textbf {\bibinfo {volume} {107}},\ \bibinfo {pages}
  {024011} (\bibinfo {year} {2023}{\natexlab{a}})},\ \Eprint
  {http://arxiv.org/abs/2209.02209} {arXiv:2209.02209 [gr-qc]} \BibitemShut
  {NoStop}%
\bibitem [{\citenamefont {Liang}\ \emph {et~al.}(2023)\citenamefont {Liang},
  \citenamefont {Xu}, \citenamefont {Mai},\ and\ \citenamefont
  {Shao}}]{Liang:2022gdk}%
  \BibitemOpen
  \bibfield  {author} {\bibinfo {author} {\bibfnamefont {D.}~\bibnamefont
  {Liang}}, \bibinfo {author} {\bibfnamefont {R.}~\bibnamefont {Xu}}, \bibinfo
  {author} {\bibfnamefont {Z.-F.}\ \bibnamefont {Mai}}, \ and\ \bibinfo
  {author} {\bibfnamefont {L.}~\bibnamefont {Shao}},\ }\href {\doibase
  10.1103/PhysRevD.107.044053} {\bibfield  {journal} {\bibinfo  {journal}
  {Phys. Rev. D}\ }\textbf {\bibinfo {volume} {107}},\ \bibinfo {pages}
  {044053} (\bibinfo {year} {2023})},\ \Eprint
  {http://arxiv.org/abs/2212.09346} {arXiv:2212.09346 [gr-qc]} \BibitemShut
  {NoStop}%
\bibitem [{\citenamefont {Xu}\ \emph {et~al.}(2023{\natexlab{b}})\citenamefont
  {Xu}, \citenamefont {Liang},\ and\ \citenamefont {Shao}}]{Xu:2023xqh}%
  \BibitemOpen
  \bibfield  {author} {\bibinfo {author} {\bibfnamefont {R.}~\bibnamefont
  {Xu}}, \bibinfo {author} {\bibfnamefont {D.}~\bibnamefont {Liang}}, \ and\
  \bibinfo {author} {\bibfnamefont {L.}~\bibnamefont {Shao}},\ }\href {\doibase
  10.3847/1538-4357/acbdfb} {\bibfield  {journal} {\bibinfo  {journal}
  {Astrophys. J.}\ }\textbf {\bibinfo {volume} {945}},\ \bibinfo {pages} {148}
  (\bibinfo {year} {2023}{\natexlab{b}})},\ \Eprint
  {http://arxiv.org/abs/2302.05671} {arXiv:2302.05671 [gr-qc]} \BibitemShut
  {NoStop}%
\bibitem [{\citenamefont {Hawking}(1976)}]{PhysRevD.13.191}%
  \BibitemOpen
  \bibfield  {author} {\bibinfo {author} {\bibfnamefont {S.~W.}\ \bibnamefont
  {Hawking}},\ }\href {\doibase 10.1103/PhysRevD.13.191} {\bibfield  {journal}
  {\bibinfo  {journal} {Phys. Rev. D}\ }\textbf {\bibinfo {volume} {13}},\
  \bibinfo {pages} {191} (\bibinfo {year} {1976})}\BibitemShut {NoStop}%
\bibitem [{\citenamefont {Bekenstein}(1973)}]{PhysRevD.7.2333}%
  \BibitemOpen
  \bibfield  {author} {\bibinfo {author} {\bibfnamefont {J.~D.}\ \bibnamefont
  {Bekenstein}},\ }\href {\doibase 10.1103/PhysRevD.7.2333} {\bibfield
  {journal} {\bibinfo  {journal} {Phys. Rev. D}\ }\textbf {\bibinfo {volume}
  {7}},\ \bibinfo {pages} {2333} (\bibinfo {year} {1973})}\BibitemShut
  {NoStop}%
\bibitem [{\citenamefont {Hawking}(1975)}]{Hawking:1975vcx}%
  \BibitemOpen
  \bibfield  {author} {\bibinfo {author} {\bibfnamefont {S.~W.}\ \bibnamefont
  {Hawking}},\ }\href {\doibase 10.1007/BF02345020} {\bibfield  {journal}
  {\bibinfo  {journal} {Commun. Math. Phys.}\ }\textbf {\bibinfo {volume}
  {43}},\ \bibinfo {pages} {199} (\bibinfo {year} {1975})},\ \bibinfo {note}
  {[Erratum: Commun.Math.Phys. 46, 206 (1976)]}\BibitemShut {NoStop}%
\bibitem [{\citenamefont {Bardeen}\ \emph {et~al.}(1973)\citenamefont
  {Bardeen}, \citenamefont {Carter},\ and\ \citenamefont
  {Hawking}}]{Bardeen:1973gs}%
  \BibitemOpen
  \bibfield  {author} {\bibinfo {author} {\bibfnamefont {J.~M.}\ \bibnamefont
  {Bardeen}}, \bibinfo {author} {\bibfnamefont {B.}~\bibnamefont {Carter}}, \
  and\ \bibinfo {author} {\bibfnamefont {S.~W.}\ \bibnamefont {Hawking}},\
  }\href {\doibase 10.1007/BF01645742} {\bibfield  {journal} {\bibinfo
  {journal} {Commun. Math. Phys.}\ }\textbf {\bibinfo {volume} {31}},\ \bibinfo
  {pages} {161} (\bibinfo {year} {1973})}\BibitemShut {NoStop}%
\bibitem [{\citenamefont {Davies}(1977)}]{Davies:1977bgr}%
  \BibitemOpen
  \bibfield  {author} {\bibinfo {author} {\bibfnamefont {P.~C.~W.}\
  \bibnamefont {Davies}},\ }\href {\doibase 10.1098/rspa.1977.0047} {\bibfield
  {journal} {\bibinfo  {journal} {Proc. Roy. Soc. Lond. A}\ }\textbf {\bibinfo
  {volume} {353}},\ \bibinfo {pages} {499} (\bibinfo {year}
  {1977})}\BibitemShut {NoStop}%
\bibitem [{\citenamefont {Geng}\ and\ \citenamefont {Lu}(2016)}]{Geng:2015kvs}%
  \BibitemOpen
  \bibfield  {author} {\bibinfo {author} {\bibfnamefont {W.-J.}\ \bibnamefont
  {Geng}}\ and\ \bibinfo {author} {\bibfnamefont {H.}~\bibnamefont {Lu}},\
  }\href {\doibase 10.1103/PhysRevD.93.044035} {\bibfield  {journal} {\bibinfo
  {journal} {Phys. Rev. D}\ }\textbf {\bibinfo {volume} {93}},\ \bibinfo
  {pages} {044035} (\bibinfo {year} {2016})},\ \Eprint
  {http://arxiv.org/abs/1511.03681} {arXiv:1511.03681 [hep-th]} \BibitemShut
  {NoStop}%
\bibitem [{\citenamefont {Heisenberg}\ \emph
  {et~al.}(2017{\natexlab{a}})\citenamefont {Heisenberg}, \citenamefont {Kase},
  \citenamefont {Minamitsuji},\ and\ \citenamefont
  {Tsujikawa}}]{Heisenberg:2017hwb}%
  \BibitemOpen
  \bibfield  {author} {\bibinfo {author} {\bibfnamefont {L.}~\bibnamefont
  {Heisenberg}}, \bibinfo {author} {\bibfnamefont {R.}~\bibnamefont {Kase}},
  \bibinfo {author} {\bibfnamefont {M.}~\bibnamefont {Minamitsuji}}, \ and\
  \bibinfo {author} {\bibfnamefont {S.}~\bibnamefont {Tsujikawa}},\ }\href
  {\doibase 10.1088/1475-7516/2017/08/024} {\bibfield  {journal} {\bibinfo
  {journal} {JCAP}\ }\textbf {\bibinfo {volume} {08}},\ \bibinfo {pages} {024}
  (\bibinfo {year} {2017}{\natexlab{a}})},\ \Eprint
  {http://arxiv.org/abs/1706.05115} {arXiv:1706.05115 [gr-qc]} \BibitemShut
  {NoStop}%
\bibitem [{\citenamefont {Heisenberg}\ \emph
  {et~al.}(2017{\natexlab{b}})\citenamefont {Heisenberg}, \citenamefont {Kase},
  \citenamefont {Minamitsuji},\ and\ \citenamefont
  {Tsujikawa}}]{Heisenberg:2017xda}%
  \BibitemOpen
  \bibfield  {author} {\bibinfo {author} {\bibfnamefont {L.}~\bibnamefont
  {Heisenberg}}, \bibinfo {author} {\bibfnamefont {R.}~\bibnamefont {Kase}},
  \bibinfo {author} {\bibfnamefont {M.}~\bibnamefont {Minamitsuji}}, \ and\
  \bibinfo {author} {\bibfnamefont {S.}~\bibnamefont {Tsujikawa}},\ }\href
  {\doibase 10.1103/PhysRevD.96.084049} {\bibfield  {journal} {\bibinfo
  {journal} {Phys. Rev. D}\ }\textbf {\bibinfo {volume} {96}},\ \bibinfo
  {pages} {084049} (\bibinfo {year} {2017}{\natexlab{b}})},\ \Eprint
  {http://arxiv.org/abs/1705.09662} {arXiv:1705.09662 [gr-qc]} \BibitemShut
  {NoStop}%
\bibitem [{\citenamefont {Babichev}\ \emph {et~al.}(2017)\citenamefont
  {Babichev}, \citenamefont {Charmousis},\ and\ \citenamefont
  {Leh\'ebel}}]{Babichev:2017guv}%
  \BibitemOpen
  \bibfield  {author} {\bibinfo {author} {\bibfnamefont {E.}~\bibnamefont
  {Babichev}}, \bibinfo {author} {\bibfnamefont {C.}~\bibnamefont
  {Charmousis}}, \ and\ \bibinfo {author} {\bibfnamefont {A.}~\bibnamefont
  {Leh\'ebel}},\ }\href {\doibase 10.1088/1475-7516/2017/04/027} {\bibfield
  {journal} {\bibinfo  {journal} {JCAP}\ }\textbf {\bibinfo {volume} {04}},\
  \bibinfo {pages} {027} (\bibinfo {year} {2017})},\ \Eprint
  {http://arxiv.org/abs/1702.01938} {arXiv:1702.01938 [gr-qc]} \BibitemShut
  {NoStop}%
\bibitem [{\citenamefont {Liu}\ \emph {et~al.}(2014)\citenamefont {Liu},
  \citenamefont {L\"u},\ and\ \citenamefont {Pope}}]{Liu:2014tra}%
  \BibitemOpen
  \bibfield  {author} {\bibinfo {author} {\bibfnamefont {H.-S.}\ \bibnamefont
  {Liu}}, \bibinfo {author} {\bibfnamefont {H.}~\bibnamefont {L\"u}}, \ and\
  \bibinfo {author} {\bibfnamefont {C.~N.}\ \bibnamefont {Pope}},\ }\href
  {\doibase 10.1007/JHEP06(2014)109} {\bibfield  {journal} {\bibinfo  {journal}
  {JHEP}\ }\textbf {\bibinfo {volume} {06}},\ \bibinfo {pages} {109} (\bibinfo
  {year} {2014})},\ \Eprint {http://arxiv.org/abs/1402.5153} {arXiv:1402.5153
  [hep-th]} \BibitemShut {NoStop}%
\bibitem [{\citenamefont {Fan}\ and\ \citenamefont {Lu}(2015)}]{Fan:2014ala}%
  \BibitemOpen
  \bibfield  {author} {\bibinfo {author} {\bibfnamefont {Z.-Y.}\ \bibnamefont
  {Fan}}\ and\ \bibinfo {author} {\bibfnamefont {H.}~\bibnamefont {Lu}},\
  }\href {\doibase 10.1103/PhysRevD.91.064009} {\bibfield  {journal} {\bibinfo
  {journal} {Phys. Rev. D}\ }\textbf {\bibinfo {volume} {91}},\ \bibinfo
  {pages} {064009} (\bibinfo {year} {2015})},\ \Eprint
  {http://arxiv.org/abs/1501.00006} {arXiv:1501.00006 [hep-th]} \BibitemShut
  {NoStop}%
\bibitem [{\citenamefont {Fan}(2016)}]{Fan:2016jnz}%
  \BibitemOpen
  \bibfield  {author} {\bibinfo {author} {\bibfnamefont {Z.-Y.}\ \bibnamefont
  {Fan}},\ }\href {\doibase 10.1007/JHEP09(2016)039} {\bibfield  {journal}
  {\bibinfo  {journal} {JHEP}\ }\textbf {\bibinfo {volume} {09}},\ \bibinfo
  {pages} {039} (\bibinfo {year} {2016})},\ \Eprint
  {http://arxiv.org/abs/1606.00684} {arXiv:1606.00684 [hep-th]} \BibitemShut
  {NoStop}%
\bibitem [{\citenamefont {Fan}(2018)}]{Fan:2017bka}%
  \BibitemOpen
  \bibfield  {author} {\bibinfo {author} {\bibfnamefont {Z.-Y.}\ \bibnamefont
  {Fan}},\ }\href {\doibase 10.1140/epjc/s10052-018-5540-7} {\bibfield
  {journal} {\bibinfo  {journal} {Eur. Phys. J. C}\ }\textbf {\bibinfo {volume}
  {78}},\ \bibinfo {pages} {65} (\bibinfo {year} {2018})},\ \Eprint
  {http://arxiv.org/abs/1709.04392} {arXiv:1709.04392 [hep-th]} \BibitemShut
  {NoStop}%
\bibitem [{\citenamefont {Wald}(1993)}]{Wald:1993nt}%
  \BibitemOpen
  \bibfield  {author} {\bibinfo {author} {\bibfnamefont {R.~M.}\ \bibnamefont
  {Wald}},\ }\href {\doibase 10.1103/PhysRevD.48.R3427} {\bibfield  {journal}
  {\bibinfo  {journal} {Phys. Rev. D}\ }\textbf {\bibinfo {volume} {48}},\
  \bibinfo {pages} {R3427} (\bibinfo {year} {1993})},\ \Eprint
  {http://arxiv.org/abs/gr-qc/9307038} {arXiv:gr-qc/9307038} \BibitemShut
  {NoStop}%
\bibitem [{\citenamefont {Iyer}\ and\ \citenamefont
  {Wald}(1994)}]{Iyer:1994ys}%
  \BibitemOpen
  \bibfield  {author} {\bibinfo {author} {\bibfnamefont {V.}~\bibnamefont
  {Iyer}}\ and\ \bibinfo {author} {\bibfnamefont {R.~M.}\ \bibnamefont
  {Wald}},\ }\href {\doibase 10.1103/PhysRevD.50.846} {\bibfield  {journal}
  {\bibinfo  {journal} {Phys. Rev. D}\ }\textbf {\bibinfo {volume} {50}},\
  \bibinfo {pages} {846} (\bibinfo {year} {1994})},\ \Eprint
  {http://arxiv.org/abs/gr-qc/9403028} {arXiv:gr-qc/9403028} \BibitemShut
  {NoStop}%
\bibitem [{\citenamefont {Casana}\ \emph {et~al.}(2018)\citenamefont {Casana},
  \citenamefont {Cavalcante}, \citenamefont {Poulis},\ and\ \citenamefont
  {Santos}}]{Casana:2017jkc}%
  \BibitemOpen
  \bibfield  {author} {\bibinfo {author} {\bibfnamefont {R.}~\bibnamefont
  {Casana}}, \bibinfo {author} {\bibfnamefont {A.}~\bibnamefont {Cavalcante}},
  \bibinfo {author} {\bibfnamefont {F.~P.}\ \bibnamefont {Poulis}}, \ and\
  \bibinfo {author} {\bibfnamefont {E.~B.}\ \bibnamefont {Santos}},\ }\href
  {\doibase 10.1103/PhysRevD.97.104001} {\bibfield  {journal} {\bibinfo
  {journal} {Phys. Rev. D}\ }\textbf {\bibinfo {volume} {97}},\ \bibinfo
  {pages} {104001} (\bibinfo {year} {2018})},\ \Eprint
  {http://arxiv.org/abs/1711.02273} {arXiv:1711.02273 [gr-qc]} \BibitemShut
  {NoStop}%
\bibitem [{\citenamefont {Smarr}(1973)}]{Smarr:1972kt}%
  \BibitemOpen
  \bibfield  {author} {\bibinfo {author} {\bibfnamefont {L.}~\bibnamefont
  {Smarr}},\ }\href {\doibase 10.1103/PhysRevLett.30.71} {\bibfield  {journal}
  {\bibinfo  {journal} {Phys. Rev. Lett.}\ }\textbf {\bibinfo {volume} {30}},\
  \bibinfo {pages} {71} (\bibinfo {year} {1973})},\ \bibinfo {note} {[Erratum:
  Phys.Rev.Lett. 30, 521--521 (1973)]}\BibitemShut {NoStop}%
\bibitem [{\citenamefont {Avramov}\ \emph {et~al.}(2023)\citenamefont
  {Avramov}, \citenamefont {Dimov}, \citenamefont {Radomirov}, \citenamefont
  {Rashkov},\ and\ \citenamefont {Vetsov}}]{Avramov:2023eif}%
  \BibitemOpen
  \bibfield  {author} {\bibinfo {author} {\bibfnamefont {V.}~\bibnamefont
  {Avramov}}, \bibinfo {author} {\bibfnamefont {H.}~\bibnamefont {Dimov}},
  \bibinfo {author} {\bibfnamefont {M.}~\bibnamefont {Radomirov}}, \bibinfo
  {author} {\bibfnamefont {R.~C.}\ \bibnamefont {Rashkov}}, \ and\ \bibinfo
  {author} {\bibfnamefont {T.}~\bibnamefont {Vetsov}},\ }\href@noop {} {\
  (\bibinfo {year} {2023})},\ \Eprint {http://arxiv.org/abs/2302.11998}
  {arXiv:2302.11998 [gr-qc]} \BibitemShut {NoStop}%
\bibitem [{\citenamefont {Traschen}(1999)}]{Traschen:1999zr}%
  \BibitemOpen
  \bibfield  {author} {\bibinfo {author} {\bibfnamefont {J.~H.}\ \bibnamefont
  {Traschen}},\ }in\ \href@noop {} {\emph {\bibinfo {booktitle} {{1999 Londrona
  Winter School on Mathematical Methods in Physics}}}}\ (\bibinfo {year}
  {1999})\ \Eprint {http://arxiv.org/abs/gr-qc/0010055} {arXiv:gr-qc/0010055}
  \BibitemShut {NoStop}%
\bibitem [{\citenamefont {Altamirano}\ \emph {et~al.}(2014)\citenamefont
  {Altamirano}, \citenamefont {Kubiznak}, \citenamefont {Mann},\ and\
  \citenamefont {Sherkatghanad}}]{Altamirano:2014tva}%
  \BibitemOpen
  \bibfield  {author} {\bibinfo {author} {\bibfnamefont {N.}~\bibnamefont
  {Altamirano}}, \bibinfo {author} {\bibfnamefont {D.}~\bibnamefont
  {Kubiznak}}, \bibinfo {author} {\bibfnamefont {R.~B.}\ \bibnamefont {Mann}},
  \ and\ \bibinfo {author} {\bibfnamefont {Z.}~\bibnamefont {Sherkatghanad}},\
  }\href {\doibase 10.3390/galaxies2010089} {\bibfield  {journal} {\bibinfo
  {journal} {Galaxies}\ }\textbf {\bibinfo {volume} {2}},\ \bibinfo {pages}
  {89} (\bibinfo {year} {2014})},\ \Eprint {http://arxiv.org/abs/1401.2586}
  {arXiv:1401.2586 [hep-th]} \BibitemShut {NoStop}%
\bibitem [{\citenamefont {Mai}\ and\ \citenamefont {Yang}(2021)}]{Mai:2020sac}%
  \BibitemOpen
  \bibfield  {author} {\bibinfo {author} {\bibfnamefont {Z.-F.}\ \bibnamefont
  {Mai}}\ and\ \bibinfo {author} {\bibfnamefont {R.-Q.}\ \bibnamefont {Yang}},\
  }\href {\doibase 10.1103/PhysRevD.104.044008} {\bibfield  {journal} {\bibinfo
   {journal} {Phys. Rev. D}\ }\textbf {\bibinfo {volume} {104}},\ \bibinfo
  {pages} {044008} (\bibinfo {year} {2021})},\ \Eprint
  {http://arxiv.org/abs/2101.00026} {arXiv:2101.00026 [gr-qc]} \BibitemShut
  {NoStop}%
\end{thebibliography}%

\end{document}